\documentclass[12pt]{article}
\usepackage{doc}
\usepackage{epsf}

\title{Coarsening of Surface Structures in Unstable Epitaxial Growth}
\author{Martin Rost\thanks{Email: marost@theo-phys.uni-essen.de.
 Fax: +49-201-183 2120.} and Joachim Krug \\
{\small Fachbereich Physik} \\
{\small Universit\"at GH Essen} \\
{\small D-45117 Essen, Germany}
}

\begin{document}
\maketitle
\begin{abstract}
We study unstable epitaxy on singular surfaces using 
continuum equations with a prescribed slope-dependent surface
current. We derive scaling relations for the late stage of
growth, where power law coarsening of the mound morphology
is observed. For the lateral size of
mounds we obtain $\xi \sim t^{1/z}$ with $z \! \geq \! 4$. 
An analytic treatment within a self--consistent mean--field approximation
predicts multiscaling of the height--height correlation function,
while the direct numerical solution of the continuum equation shows 
conventional scaling with $z=4$, independent of the shape of the 
surface current. 
\end{abstract}

\section{Introduction}
On many crystal surfaces step edge barriers are observed which prevent
interlayer (downward) hopping of diffusing adatoms
\cite{Ehrlich,Schwoebel}. In homoepitaxy from a molecular beam this
leads to a growth instability  which can be understood on a basic
level: Adatoms form islands on the
initial substrate and matter deposited on top of them is caught there
by the step edge barrier. Thus a pyramid structure of islands 
on top of islands develops.
 
At late stages of growth pyramids coalesce and form large
``mounds''. Their lateral size $\xi$ is found 
experimentally to increase according to
a power law in time, $\xi \sim t^{1/z}$ with $z \simeq 2.5$ -- 6
depending on the material and, possibly, deposition conditions used.
A  second characteristic is
the slope of the mounds' hillsides $s$, which is observed to 
either approach a constant (often referred to as
a ``magic slope'' since it does not necessarily coincide with a
high symmetry plane) or
to increase with time as $s \sim t^{\alpha}$ \cite{exps,advances}. The
surface width (or the height of the mounds) then grows as $w \sim s \xi
\sim t^\beta$
with $\beta = 1/z + \alpha$, where $\alpha = 0$ for the case of magic
slopes.

On a macroscopic level these instabilities can be understood in terms of a 
growth-induced, slope-dependent  surface current \cite{Villain,KPS}. 
Since diffusing
adatoms preferably attach to steps from the terrace {\em below},
rather than from
{\em above}, the current is uphill and destabilizing. 
The concentration of diffusing adatoms 
is maintained by the incoming particle flux; thus, the surface current
is a nonequilibrium effect.

The macroscopic view is quantified in a continuum growth equation,
which has been proposed and studied by several groups
\cite{Johnson,Stroscio,Siegert1,Siegert2,Sander,masch,paveljoachim}.
The goal of the present contribution is to obtain analytic estimates
for the scaling exponents and scaling functions of this continuum theory.
To give an outline of the article: In the next section we briefly
introduce the continuum equations of interest. 
A simple scaling ansatz, presented in
Section 3, leads to scaling relations 
and inequalities for the exponents $1 \! / \! z,
\alpha$ and $\beta$.
In Section 4 we present a solvable mean--field model for the dynamics
of the height--height correlation
function. Up to logarithmic corrections, 
the relations of Section 3 are corroborated. Finally, in the concluding
Section 5 the mean--field correlation functions are compared to
numerical simulations of the full growth equation, 
and the special character of the mean--field
approximation is pointed out.

\section{Continuum equation for MBE}
Under conditions typical of molecular beam epitaxy (MBE), evaporation and
the formation of bulk defects can be neglected. 
The height 
$H({\bf x},t)$ of the surface above the substrate plane
then satisfies a continuity equation, 
\begin{equation}
\label{cont1}
\partial_t H + \; \; \nabla \! \! \cdot \! \! {\bf J}_{\mbox{surface}}\{ H \} = F,
\end{equation}
where $F$ is the incident mass flux out of the molecular beam. Since we are
interested in large scale features we neglect fluctuations in $F$
(``shot noise'')  and in the surface current (``diffusion
noise''). In general, the systematic current
${\bf J}_{\mbox{surface}}$ depends on the whole surface
configuration. Keeping only the most important 
terms in a gradient 
expansion\footnote{We follow the common practice and disregard contributions
to the current which are {\em even} in $h$,
such as $\nabla (\nabla h)^2$, though they
may well be relevant for the coarsening behavior of the surface
\cite{Stroscio,Politi}.}, subtracting the mean height 
$H \! = \! Ft$, and using appropriately rescaled units of height, 
distance and time \cite{paveljoachim}, Eq.\ (\ref{cont1}) attains the
dimensionless form
\begin{equation}
\label{cont2}
\partial_t h = - (\nabla^2)^2 h - \nabla \cdot f(\nabla h^2) \; \nabla h.
\end{equation}
The linear term describes relaxation through adatom diffusion driven by the surface free
energy \cite{Mullins}, while the second nonlinear term models the
nonequilibrium current \cite{Villain,KPS}. Assuming in-plane symmetry,
it follows that the
nonequilibrium current is (anti)parallel to the local tilt $\nabla h$,
with a magnitude $f(\nabla h^2)$ depending only on the magnitude of the tilt.
We consider two different forms for the function $f(\nabla h^2)$:

\noindent
(i) Within a Burton-Cabrera-Frank-type theory \cite{advances,masch,Politi}, 
for small tilts the current is
proportional to $| \nabla h |$, and in the opposite limit it is
proportional to $| \nabla h|^{-1}$. This suggests the interpolation
formula \cite{Johnson} $f(s^2) = 1/(1+s^2)$. Since we are interested in 
probing the dependence on the asymptotic decay of the current for large
slopes, we consider the generalization
\begin{equation}
\label{model(i)}
f(s^2) = 1/( 1 + |s|^{1 + \gamma}) \;\;\; {\rm [model \; (i)]}.
\end{equation}
Since $\gamma = 1$ also in the extreme case of complete suppression of interlayer transport
\cite{masch,wedding}, physically reasonable values of $\gamma$ are restricted to 
$\gamma \geq 1$.

\noindent
(ii) Magic slopes can be incorporated into the continuum description by letting the nonequilibrium
current change sign at some nonzero tilt \cite{KPS,Siegert1,Siegert2}. 
A simple choice, which places the magic slope at $s^2=1$, is
\begin{equation}
\label{model(ii)}
f(s^2) = 1 - s^2 \;\;\; {\rm [model \; (ii)]};
\end{equation}
a microscopic calculation of the surface current for a model
exhibiting magic slopes has been reported by Amar and Family \cite{amar}.

The stability properties of a surface with uniform slope ${\bf m}$
are obtained by inserting the ansatz
$h({\bf x},t) = {\bf m \! \cdot \! x} + \epsilon({\bf x},t)$ 
into (\ref{cont2}) and expanding to linear order in $\epsilon$. One obtains
\begin{equation}
  \label{linstab}
  \partial_t \epsilon = \bigl[ \nu_\| \partial_\|^2 + \nu_\perp
  \partial_\perp^2 - (\nabla^2)^2 \bigr] \epsilon, 
\end{equation}
where $\partial_\| (\partial_\perp)$ denotes the partial derivative parallel
(perpendicular) to the tilt {\bf m}. The coefficients are
$\nu_\| = -(d/d|{\bf m}|) |{\bf m}|f({\bf m}^2)$ and $\nu_\perp = -
f({\bf m}^2)$. If one of them is negative, the surface is unstable to
fluctuations varying in the corresponding direction: Variations
perpendicular to {\bf m} will grow when the current is uphill (when $f > 0$), while
variations in the direction of {\bf m} grow when the current is an 
increasing function of the tilt. Both models have a change in the sign
of $\nu_\|$, model (i) at $|{\bf m}| = \gamma^{-1/(1+\gamma)}$, model
(ii) at $|{\bf m}| = 1/\sqrt{3}$. For model (i) $\nu_\perp < 0$ always,
corresponding to the step meandering instability of Bales and Zangwill 
\cite{paveljoachim,Bales}. In contrast, for model (ii) the current is downhill for
slopes $|{\bf m}| > 1$, and these surfaces are absolutely stable. 

In this work we focus on singular surfaces, ${\bf m} \!
= \! 0$, which are unstable in both models; coarsening behavior
of vicinal surfaces has been studied elsewhere \cite{paveljoachim}. 
The situation envisioned in the rest of this article is the following:
For solutions of the PDE (\ref{cont2}) we choose a
flat surface with small random fluctuations $\epsilon({\bf x})$ as initial
condition.
Mostly the initial fluctuations will be uncorrelated in space, though
the effect of long range initial correlations is briefly addressed in
Section \ref{meanfield}. The fluctuations are amplified by the linear
instability, and eventually the surface enters the late time coarsening
regime that we wish to investigate.

\section{Scaling Relations and Exponent Inequalities}
\label{scalingansatz}
In this section we assume
that in the late time regime the solution of
(\ref{cont2}) is described by a scaling form, namely that
the surface $h({\bf x},t)$ at time $t$ has the same (statistical)
properties as the rescaled surface $\tau^{-\beta} \; h({\bf
  x}/\tau^{1/z},\tau t)$ at time $\tau t$. The equal time height--height
correlation function $G({\bf x},t) \equiv \langle h({\bf x},t) \;
h(0,t) \rangle$ then has a scaling form 
\begin{equation}
G({\bf x},t) = w(t)^2 g(|{\bf x}|/\xi(t)), 
\end{equation}
where the relevant lengthscales are the surface width
$w(t) \! = \! \langle h({\bf x},t)^2 \rangle^{1/2} \! \sim \!
t^{\beta}$, i.e.\ the typical height of the mounds, and their lateral
size $\xi(t) \! \sim \! t^{1/z}$, given by the first zero of $G$. These
choices correspond to $g(0) \! = \! 1$ and $g(1) \! = \! 0$. Moreover
they lead to a definition of the  typical slope of mounds as $ s \equiv
w/\xi \sim t^\alpha $ with $\alpha = \beta - 1/z$.

We start our reasoning with the time dependence of the width
\begin{equation}
\label{widthdyn}
\frac{1}{2} \partial_t w^2(t) = - \langle \left( \Delta h({\bf x},t)
\right)^2 \rangle + \langle \nabla h({\bf x},t)^2 
f( \nabla h({\bf x},t)^2 ) \rangle \equiv - I_1 + I_2. 
\end{equation}
Clearly $I_1 \geq 0$. Since we expect the width to increase with time,
we obtain
the inequalities
\begin{equation}
\label{ineqa}
0 \leq \frac{1}{2} \partial_t w^2(t) \leq I_2
\end{equation}
and 
\begin{equation}
\label{ineqb}
I_1 \leq I_2.
\end{equation}

The first conclusion can be drawn even without the scaling assumption:
For model (ii) and model (i) with $\gamma \geq 1$,
$(\nabla h)^2 f(\nabla h^2)$ has an upper bound, and so has $I_2$. 
Therefore $\partial_t w^2 \leq const$. We conclude that
the increase of the width $w(t)$ cannot be faster than $t^{1/2}$ if it
is caused by a destabilizing nonequilibrium current on a surface with
step edge barriers.

Assuming scaling we estimate $I_1 \sim (s/\xi)^2$ and 
$\partial_t w^2 \sim w^2/t \sim (s \xi)^2/t$. For model (i) we
further have $I_2 \sim s^2 f(s^2) \sim s^{1-\gamma}$. 
In terms of the scaling
exponents $\alpha$ and $1/z$ inequality (\ref{ineqa}) yields $2(\alpha + 1/z) -1
\leq \alpha (1 - \gamma)$, while the second inequality (\ref{ineqb}) leads to $2
\alpha - 2/z \leq \alpha ( 1 - \gamma)$. Combining both inequalities we have
\begin{equation}
  \label{ineq1}
  \frac{1 + \gamma}{2} \; \alpha \; \; \; \leq \; \; \; \frac{1}{z} \;
  \; \; \leq \; \; \; \frac{1}{2} - 
  \frac{1 + \gamma}{2} \; \alpha.
\end{equation}

To proceed we note that an upper bound on the lateral mound size
$\xi$ can be obtained from the requirement that the mounds should be 
stable against the Bales-Zangwill step meandering instability 
\cite{paveljoachim,Bales}: Otherwise they would break up into smaller mounds.
>From (\ref{linstab}) it is easy to see that, for the large slopes of interest here,
fluctuations of a wavelength exceeding $2 \pi / \sqrt{|\nu_\perp |}$ are unstable.
Since $- \nu_\perp = f(s^2) \sim s^{-(1+\gamma)}$, we impose the condition
$\xi
\leq 2 \pi / \sqrt{|\nu_\perp |} \sim m^{(1 + \gamma)/2}$ or, in terms of
scaling exponents,
\begin{equation}
  \label{ineq2}
  \frac{1}{z} \; \; \; \leq \; \; \; \frac{1 + \gamma}{2} \; \alpha.
\end{equation}
Hence the first relation in (\ref{ineq1}) becomes an equality
(which was previously derived for the one-dimensional case \cite{advances}), and the
second relation yields
\begin{equation}
  \label{scalineq}
  z \geq 4, \; \; \; \alpha \; = \; \frac{2}{z(1 + \gamma)} 
  \; \leq \; \frac{1}{2(1 + \gamma)}, \;
  \; \; \beta \; \leq \; \frac{3 + \gamma}{4(1 + \gamma)}.
\;\;\; [{\rm model \; (i)}]
\end{equation}

For model (ii) we assume that the slope $s$ approaches its stable
value $s=1$ 
as $s \sim 1 - t^{-\alpha'}$ with $\alpha' > 0$. 
The estimate of the last term in (\ref{widthdyn}) then becomes
$I_2 \sim s^2 (1-s^2) \simeq 1 - s^2 \sim t^{-\alpha'}$. Thus
inequality (\ref{ineqa}) yields $2/z - 1 \leq - \alpha'$, and from 
(\ref{ineqb}) it follows follows that
$-2/z \leq - \alpha'$. As for model (i) the next estimation uses $\xi
\leq 2 \pi/\sqrt{|\nu_\perp|}$ with $|\nu_\perp| = 1 - s^2 \sim
t^{-\alpha'}$. Again we obtain the inverse of the second of the
above inequalities, {\em viz.} $1/z
\leq \alpha' /2$. Altogether this yields
\begin{equation}
  \label{scalineq2}
  \frac{1}{z} = \frac{\alpha'}{2} = \beta \leq \frac{1}{4}. 
\;\;\; [{\rm model \;  (ii)}]
\end{equation}

To summarize the general results obtained in this section: 
In addition to the bound on the temporal increase of 
the surface width,
$w(t) < const.
\; t^{1/2}$, the scaling ansatz yields an upper bound on the increase
of the lateral length scale, $\xi(t) < const. \; t^{1/4}$, valid for
both models. A more
elaborate approximation, to be presented in the next section,
predicts the above
inequalitites (\ref{scalineq},\ref{scalineq2}) to hold as equalities
(up to logarithmic corrections).

\section{Spherical Approximation}
\label{meanfield}

We consider the time dependence of the equal time height-height
correlation function defined above:
\begin{equation}
\label{corrdyn}
\partial_t G({\bf x},t) = - 2 \; \Delta^2 G({\bf x},t) - 2 \; \nabla \cdot \langle
h(0,t) \; f(\nabla h({\bf x},t)^2) \; \nabla h({\bf x},t) \rangle,
\end{equation}
where $\Delta = \nabla^2$ is the Laplace operator.
In order to obtain a closed equation for $G({\bf x},t)$ we replace
$f(\nabla h^2)$ by $f(\langle \nabla h^2 \rangle)$ in the second term
on the right hand side. This approach is inspired by the spherical
`large $n$' limit of phase ordering kinetics \cite{bray}, and will
be referred to as the spherical approximation. The 
argument of $f$ is then easily expressed in terms of $G$:
\begin{equation}
\langle \nabla h({\bf x},t)^2 \rangle = - \Delta \langle h(x,t)^2
\rangle =  - \Delta G(0,t),
\end{equation}
and the closure of (\ref{corrdyn}) reads
\begin{equation}
\label{cdmf}
\partial_t G({\bf x},t) = - 2 \; \Delta^2 G({\bf x},t) - 2 \; f(|\Delta G(0,t)|) \;
\Delta G({\bf x},t).
\end{equation}
Since we consider dynamics which are isotropic in substrate space, and
also isotropic distributions of initial conditions, $G({\bf x},t)$ will only
depend on $|{\bf x}|$ and $t$. Consequently we consider the structure factor
$S(k,t)$ as a function of  $k = |{\bf k}|$ and $t$, which satisfies
\begin{equation}
\label{Seq}
\partial_t S(k,t) = -2 \left[ k^4 - f(a(t)) k^2 \right] S(k,t).
\end{equation}
Here we have defined the function $a(t)$ through
\begin{equation}
\label{a(t)}
a(t) =  (2\pi^{d/2}/\Gamma(d/2)) 
\! \int_0^{\infty} \! \! dk \; k^{d+1} S(k,t),
\end{equation}
and $d$ denotes the surface dimensionality ($d=2$ for real surfaces). 
The formal solution of (\ref{Seq}) then reads
\begin{equation}
\label{formal}
S(k,t) = S_0(k) \exp \left[ -2 t k^4 + 2 k^2 \! \int_0^t \! \! ds \; f(a(s)) \right].
\end{equation}
The initial condition $S_0(k)$ reflects the disorder in the initial
configuration of Eq.\ (\ref{cont1}). It consists of fluctuations at
early times, i.e.\ the first nucleated islands, from which mounds
will later develop. Simulations of
microscopic models for MBE on singular surfaces at submonolayer coverages 
\cite{harald} indicate the following shape
of $S_0(k)$: From a hump at some finite wavenumber, corresponding to
the typical distance $\ell_D$ between island nuclei, it falls off to
zero for $k \to \infty$. For $k \to 0$ it goes down to a {\em finite}
value $c > 0$.

At late times the hump in $S(k,t)$ persists, situated at some
$k_{max}$ near the maximum of the exponential in (\ref{formal}). 
It belongs to a lateral lengthscale $\xi$, denoting the typical
distance of neighboring mounds. For
late times $k_{max}$ will go to zero, so we need only consider $S_0(k)$
near $k = 0$. In fact for the leading contribution to $k_{max}$
(and the leading power in $\xi$) we only need $S_0(k) \equiv c$.
More detailed remarks on the case $\lim_{k \to 0}S_0(k)=0$ and on the
presence of long range correlations in the initial stage
can be found at the end of this section. The particular value of $c$
has no influence on
the coarsening exponent, so we take $S_0(k) = (2 \pi)^{-d/2}$ which
corresponds to $G({\bf x},t \! = \! 0) = \delta({\bf x})$.

To follow the analysis, note that $a(t)$ is a functional (\ref{a(t)}) of 
$S(k,t)$, and on the other hand it is used for the calculation
of $S(k,t)$. This imposes a condition of self consistency on the
solution, which we write as follows
\begin{equation}
\label{selfc}
\frac{db}{dt} = f \left( 2/(2^{d/2}\Gamma(d/2)) 
\int_0^{\infty} \! \! dk \; k^{d+1} \exp \left(-2tk^4 +
2k^2 b(t) \right) \right).
\end{equation}
We used the initial conditions motivated above, and
the shorthand $b(t) \! = \! \int_0^t ds \; f(a(s))$. The integral can be
evaluated,
yielding for $b(t)$ the differential equation
\begin{equation}
\label{dgl}
\frac{db}{dt} = f \left( \frac{d}{2} (4t)^{-(d+2)/4} \; 2^{-d/2} \; 
D_{ \! -\frac{d+2}{2}} ( -b/\sqrt{t}) \; 
\exp \frac{b^2}{4t} \right),
\end{equation}
where $D$ denotes a parabolic cylinder function \cite{gradstein}.
Equation (\ref{dgl}) cannot
be solved explicitly, but for the late time behavior we can use
an asymptotic approximation for $D$,
since its argument $b/\sqrt{t} \to \infty$ for $t \to \infty$. To see this,
note that (\ref{dgl}) is of the form
\begin{equation}
\frac{db}{dt} = f \left( t^{-(d+2)/4} 
\; F \! \left( \! \frac{b}{\sqrt{t}} \! \right) \right).
\end{equation}
Therefore, if $b/\sqrt{t}$ remained bounded, for large $t$ the argument of $f$
in Eq.\ (\ref{dgl}) would be close to 0, and (\ref{dgl}) would approximately
be $db/dt \simeq f(0) = 1$. This is in contradiction to the assumption
$b <$ const.$\times \sqrt{t}$ which therefore cannot be true.

For large $t$ (and large $b/\sqrt{t}$) we then approximate (\ref{dgl}) by
\begin{equation}
\frac{db}{dt} = f \left( \sqrt{\frac{\pi}{2^{d-1}}} \; 
(4t)^{-(d+2)/4} /\Gamma(d/2) \;
\left( \frac{b}{\sqrt{t}} \right)^{d/2} \exp \frac{b^2}{2t} \right).
\end{equation}
This shows that $b/\sqrt{t}$ must grow more slowly than
any power of $t$: If $b \sim t^{1/2 + \epsilon}$ then $db/dt \sim 
t^{\epsilon - 1/2}$, whereas the argument in $f$ would increase
exponentially, dominated by a term $\exp t^{2 \epsilon}$. For both
choices (\ref{model(i)}) and (\ref{model(ii)}) of $f$ the right hand side of
(\ref{dgl}) would decrease much faster than the left or even become negative.

Depending on the choice of the current function $f$ we get different asymptotic
behaviors in the leading logarithmic increase of $B = b^2/(2t)$.
We first consider the case (ii), where $f(s^2) = 1 - s^2$. Here Eq.\ (\ref{dgl}) reads
\begin{equation}
t \; \frac{d B}{dt} = - B + \sqrt{2 B t}
\left( 1 - C_{\rm (ii)} \; B^{d/4} \; t^{-(d+2)/4} \; \exp B \right),
\end{equation}
where $C_{\rm (ii)}$ is a constant without any interest.
None of the terms must increase with time as a power of $t$. Hence asymptotically
the term in brackets must vanish, which requires that $\exp B \sim t^{(d+2)/4}$.
The leading behavior of $B$ is therefore
\begin{equation}
\label{betaii}
B \simeq \frac {d+2}{4} \; \log t.
\end{equation}
Similarly we treat case (i), using the asymptotic behavior
$f(s^2) \simeq (s^2)^{-\gamma/2}$ for large $s^2$. Equation (\ref{dgl}) then becomes
\begin{equation}
t \; \frac{d B}{dt} = - B + C_{\rm (i)} \; t^{(\gamma/2)(d+2)/4 + 1/2}
\; B^{-(\gamma/2)d/4 + 1/2} \exp -(\gamma/2) B.
\end{equation}
Again the powers of $t$ in the last term must cancel, yielding
\begin{equation}
\label{betaiii}
B \simeq \left( \frac {d+2}{4} + \frac{1}{\gamma} \right) \; \log t.
\end{equation}
There is a noteworthy correspondence between models (i) and (ii): The
solution of (ii) is the limit $\gamma \to \infty$ of the solution of (i). 
In this sense,
a current which 
vanishes at a finite slope is equivalent to a positive shape function $f(s^2)$
decreasing faster than any power of $s$. The same correspondence
applies also on the level of the inequalities derived in Section 
\ref{scalingansatz}, as can be seen by letting $\gamma \to \infty$
in (\ref{scalineq}) and comparing to (\ref{scalineq2}). 

The asymptotic form of $b(t)$ gives us the following time
dependence of the coarsening surface structure: Inserting $b(t)$ into
the expression for the structure factor $S(k,t)$ (\ref{formal}) we obtain
for each time $t$ a wavenumber
\begin{equation}
\label{km}
k_{m}(t) = \left( \frac{1}{2} \left( \frac{d+2}{8} + \frac{1}{\gamma}
  \right) \; \; \frac{\log t}{t} \right)^{1/4},
\end{equation}
which has the maximal contribution to $S(k,t)$. It can be interpreted as the
inverse of a typical lateral lengthscale $\xi \sim (t/\log t)^{1/4}$.
Up to a logarithmic factor, 
we obtain lateral coarsening with a power $1/4$ for both choices of
$f(s^2)$. This corresponds to $z=4$, which saturates the bound derived in 
Section \ref{scalingansatz}. 

It is however important to note that the resulting structure factor
{\em cannot} be written in a simple scaling form 
$S(k,t) = w^2 k_m^{-d} {\cal S}(k/k_m)$, as would
be expected if $k_m^{-1}$ were the only scale in the problem \cite{bray}. Rather,
one obtains the
{\em multiscaling form} \cite{coniglio}
\begin{equation}
\label{multi}
S(k,t) = L(t)^{\varphi(k/k_m \! (t))},
\end{equation}
where $\varphi(x) = 2x^2 - x^4$, and $L(t) \sim t^{(\frac{d+2}{4} +
\frac{1}{\gamma})/d}$ is a
second lengthscale in the system. In contrast to $k_m^{-1}$, the exponent
describing the temporal increase of $L(t)$ {\em does} depend on the shape
of the current function $f$.

Next we discuss the behavior of the
typical slope of the
coarsening mounds, given by $a(t) = \langle (\nabla h)^2 \rangle$.
This is obtained directly from (\ref{selfc}).
For model (ii) $a(t)$ approaches the stable value (``magic slope'')
$s^2 \! = \! 1$, with a leading correction
\begin{equation}
\label{magic}
a(t) = 1 - \dot b(t) \simeq 1 - \frac{1}{2} \sqrt{\frac{d+2}{2}}
\left( \frac{\log t}{t} \right)^{1/2}.
\end{equation}
Note that the approach to the magic slope is very slow, 
a possible explanation for the common difficulty of deciding 
whether $a(t)$
attains a final value or grows indefinitely in numerical simulations \cite{pavel}.
We further remark that, up to a logarithmic factor, the inequality $\alpha' \leq 1/2$
derived in (\ref{scalineq2}) for the exponent describing the approach to the magic
slope becomes an equality within the spherical approximation.

For model (i) the typical slopes diverge as
\begin{equation}
\label{steep}
a(t) \simeq \dot b^{-2/(1+\gamma)} \simeq 
\left( \frac{8}{\frac{d+2}{4}+\frac{1}{\gamma}} \right)^{1/(1+\gamma)}
\left( \frac{t}{\log t} \right)^{1/(1+\gamma)},
\end{equation}
consistent with the value $\alpha = 1/(2 + 2 \gamma)$ derived as a bound
in (\ref{scalineq}). 
In the limit $\gamma \to \infty$ the slope does not increase at all,
which again is comparable to the presence of a stable slope.

To close this section we briefly comment on the
the shape of the structure factor
$S(k,t)$ and the correlation function $G({\bf x},t)$ obtained within the
spherical approximation. Assuming
initial correlations as used above, $S_0(k) \equiv c$, the structure
factor is analytical at any time $t$, as can be seen in equation
(\ref{formal}). The corresponding correlation function therefore
decays faster than any power of $|{\bf x}|$, modulated with oscillations
of wavenumber $k_{max}(t)$.

We can also predict the further evolution of long range correlations,
assuming that they were initially present. A power law decay of $G({\bf x},t \!
= \!0)$ corresponds to a singularity in $S_0(k)$. Suppose the singularity is located at
some point $k_0 > 0$ (the power decay of $G$ is then modulated by
oscillations). Then the singularity will remain present in $S(k,t)$,
but it will be suppressed as $\exp -t k_0^4$ for late times. This implies that $G({\bf x},t)$ 
has a very weak power law tail for very large $|{\bf x}|$, but
up to some $x_0$ (which increases with time) it decays faster than any power. 
However a singularity in
$S_0(k)$ will not be suppressed if it lies at the origin $k_0 = 0$, since then in Eq.\
(\ref{formal}) it is multiplied by unity. In real space, this implies that a power law
decay of correlations without oscillations will remain
present. 

Even if $S_0(k)$ is singular at $k=0$, the scaling laws derived above remain valid.
Suppose for example that $S_0(k) \sim k^{\sigma}$ for $k \to 0$.
In transforming (\ref{formal}) back to real space, such a power law singularity
can be absorbed into the phase space factor $k^{d-1}$ involved in the ${\bf k}$-integration.
The result is simply a shift in the dimensionality, $d \to d + \sigma$, which affects
the prefactors of the scaling laws (\ref{km}), (\ref{magic}) and
(\ref{steep}) for $k_m(t)$ and $a(t)$ but not the powers of 
$t/\log t$.  

\section{Conclusions}
We have presented two approximate ways to predict the late stage of
mound coarsening in homoepitaxial growth. To our knowledge this is
the first theoretical calculation of coarsening exponents for this problem.

Although we have made heavy use of concepts developed in phase--ordering
kinetics \cite{bray}, our results cannot be directly inferred from the existing
theories in that field.
As was explained in detail by Siegert
\cite{Siegert2}, equation (\ref{cont2}) rewritten
for the slope ${\bf u} \equiv \nabla h$ 
has the form of a relaxation dynamics
driven by a
generalized free energy, $\dot {\bf u} = \nabla \nabla \cdot \delta
{\cal F}({\bf u})/ \delta {\bf u}$. Phase ordering with a conserved vector
order parameter ${\bf m}$ is described by a similar form, $\dot {\bf m} = \nabla
\cdot \nabla \delta {\cal F}({\bf m})/ \delta {\bf m}$, however it appears that the
interchange of the order of the differential operators,
from $\nabla^2$ to $\nabla \nabla \cdot$, may lead to a
qualitatively different behavior \cite{Siegert2}.

Nevertheless, the results obtained so far must be refined. Ideally, one would
like to derive {\em equalities} for the exponents using the scaling
ansatz of Section \ref{scalingansatz}. More modestly,
it would be desirable to extend the approach so that the effects of
current functions without in-plane isotropy \cite{Siegert1,Siegert2} and
of contributions proportional to $\nabla (\nabla h)^2$ \cite{Stroscio,Politi}
on the scaling behavior can be assessed.

The main drawback of the spherical approximation in Section \ref{meanfield} is
that it does
not predict conventional scaling. The experience from phase--ordering kinetics in the
$O(n)$ model suggests that the multiscaling behavior
obtained above may be an artefact of the spherical approximation \cite{bray,coniglio}.
To address this issue, we have carried out a 
numerical integration of Equation (\ref{cont2}), with weak
uncorrelated noise as initial condition. The results indicate 
conventional scaling behavior in the late stage of growth, with
exponents $z=4$, $\alpha = 1/(2 + 2 \gamma)$, which saturate the
bounds of Section \ref{scalingansatz}. 

\begin{figure}[htbp]
  \begin{centering}
    \vspace*{5cm}
    \epsfxsize 12cm
    \epsfbox{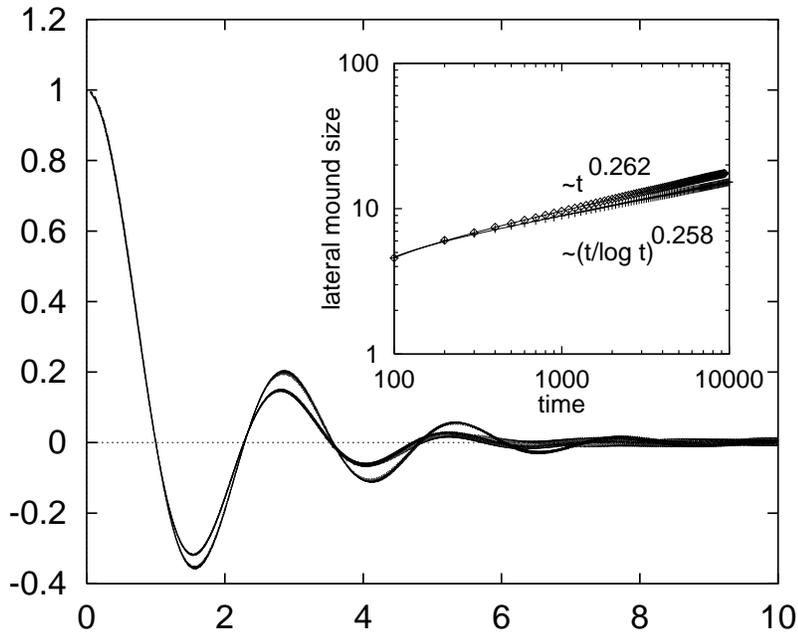}
    \caption{Comparison of numerical integration of Eq.\ (\protect
    \ref{cont1}) and the spherical approximation described in Section
    \protect \ref{meanfield}, both for model (i) with $\gamma \! = \!
    1$. Main figure shows a scaling plot of the correlation 
    function $G(|{\bf x}|,t)$ 
    with the first zero and $G(0,t)$ rescaled to unity. For large
    $|{\bf x}|$ the approximate solution differs from the full
    model by more pronounced oscillations. The inset
    shows the evolution of the first zero of $G$, also indicating best
    fits to the form  $t^{1/z}$ (full dynamics) and $(t/\log
    t)^{1/z}$ (spherical approximation).}
    \label{fig:scalplot}
  \end{centering}
\end{figure}

Figure \ref{fig:scalplot} shows a scaling
plot of $G(|{\bf x}|,t)$ of model (i) with $\gamma = 1$ for times $t = 500, 600, \dots, 10000$
obtained from the numerical integration of (\ref{cont2}). It is
compared at
times $t = 1000, 1100, \dots, 10000$ to the spherical approximation. The
first zero and the width at $|{\bf x}| \! = \! 0$ are rescaled to
$1$. Initial conditions of the approximation were chosen to coincide
with the full dynamics at $t \! = \! 100$. The spherical approximation of $G$ takes a
slightly different shape -- its oscillations are more pronounced for
larger $|{\bf x}|$. We obtained a similar scaling plot for $\gamma \!
= \! 3$ and for model (ii).

The inset shows the evolution of the first zero of $G$: In the full
dynamics it is best 
approximated by a power law $\xi \sim \tau^{1/z}$ with $z \! = \! 3.85$
for the full dynamics, where $\tau \! = \! t \! - \! t_0$. The
spherical approximation deviates from a 
power law. Here for late times the best fit is $\xi \sim (\tau / \log
\tau)^{1/z}$ with $z \! = \! 3.87$. The beginning of the time
integration $t \! = \! 0$ does not 
coincide with the exptrapolated zero of the power laws $t_0$, because the
mounds take a finite time to develop out of the initial growth
instability. This introduces the additional fitting parameter $t_0$. The
steepening of the mounds (not shown in the graph) develops with the
power $\alpha \! = \! 0.26$. For $\gamma \! = \! 3$ in model (i) we
obtain $z \! = \! 4.18$ and $\alpha \! = \! 0.126$. For model (ii) we
refer to the intergrations of Siegert \cite{Siegert2}, which indicate
$z \! = \! 4$.

Note that the multiscaling behavior of $G$ in the spherical
approximation is very weak, in the sense that the curves at different
times do not differ in shape too much.  A more sensitive test of the 
scaling behavior of Equation (\ref{cont1}), in order to pin down the
difference to the spherical approximation, would be desirable and can
be achieved by extracting the function $\varphi$ (see Eq.\
(\ref{multi})) from the data of the numerical
integration. Conventional scaling yields $\varphi \! \equiv \!
const$. Work in this direction
is currently in progress \cite{claudio}.

\noindent
{\bf Acknowledgements} We thank P.\ \v{S}milauer, F.\ Rojas In\'\i{}guez,
A.\ J.\ Bray and C.\ Castellano for many helpful hints and discussions.
This work was supported by DFG within SFB 237 {\em Unordnung und
grosse Fluktuationen}.


\begin{thebibliography}{99}
\bibitem{Ehrlich}
G.\ Ehrlich and F.G.\ Hudda, J.\ Chem.\ Phys.\ {\bf 44}, 1039 (1966).
\bibitem{Schwoebel} R.L.\ Schwoebel and E.J.\ Shipsey, J.\ Appl.\ Phys.\
{\bf 37}, 3682 (1966); R.L.\ Schwoebel, J.\ Appl.\ Phys.\ {\bf 40}, 614 (1969).
\bibitem{exps}
Some selected experiments are
H.-J. Ernst, F. Fabre, R. Folkerts and J. Lapujoulade,  
{\em Phys. Rev. Lett.} 72:112 (1994);
C. Orme, M.D. Johnson, K.-T. Leung, B.G. Orr, P. \v{S}milauer and D. Vvedensky,
{\em J. Cryst. Growth} 150:128 (1995);
J.E. Van Nostrand, S.J. Chey, M.-A. Hasan, D.G. Cahill and J.E. Greene,
{\em Phys. Rev. Lett.} 74:1127 (1995);
K. Th\"urmer, R. Koch, M. Weber and K.H. Rieder, 
{\em Phys. Rev. Lett.} 75:1767 (1995); as well as 
Refs.\cite{Johnson,Stroscio}. 
\bibitem{advances}
A review is given in J.\ Krug, Origins of scale invariance in growth processes,
{\em Advances in Physics} (in press). 
\bibitem{Villain} J.\ Villain, J.\ de Physique I {\bf 1},1 (1991).
\bibitem{KPS}
J.\ Krug, M.\ Plischke and M.\ Siegert, Phys.\ Rev.\ Lett.\ {\bf 70}, 3271 (1993).
\bibitem{Johnson}
M.D.\ Johnson, C.\ Orme, A.W.\ Hunt, D.\ Graff, J.\ Sudijono, L.M.\ Sander and B.G.\ Orr,
Phys.\ Rev.\ Lett.\ {\bf 72}, 116 (1994).
\bibitem{Stroscio}
J.A.\ Stroscio, D.T.\ Pierce, M.\ Stiles, A.\ Zangwill and L.M.\ Sander,
Phys.\ Rev.\ Lett.\ {\bf 75}, 4246 (1995).
\bibitem{Siegert1}
M.\ Siegert and M.\ Plischke, Phys.\ Rev.\ Lett.\ {\bf 73}, 1517 (1994). 
\bibitem{Siegert2}
M.\ Siegert, 
{\em in:} ``Scale Invariance, Interfaces and Non-Equilibrium 
Dynamics", A.J.\ McKane, M.\ Droz, J.\ Vannimenus and D.E.\ Wolf, eds., Plenum Press, New York (1995), p.165.
\bibitem{Sander}
A.W.\ Hunt, C.\ Orme, D.R.M.\ Williams, B.G.\ Orr and L.M.\ Sander,
Europhys.\ Lett.\ {\bf 27}, 611 (1994).
\bibitem{masch} J. Krug and M. Schimschak, J.\ Phys.\ I France {\bf 5},
1065 (1995).
\bibitem{paveljoachim}
M.\ Rost, J.\ Krug and P.\ \v{S}milauer, Surface Science, in press.
\bibitem{Politi} P.\ Politi and J.\ Villain, Phys. Rev. {\bf 54}, 5114 (1996).
\bibitem{Mullins} W.W.\ Mullins, J.\ Appl.\ Phys.\ {\bf 30}, 77 (1959).
\bibitem{wedding} J. Krug, On the shape of wedding cakes 
(to appear in Journal of Statistical Physics).
\bibitem{amar} J.G. Amar and F. Family, Phys. Rev. B (in press).
\bibitem{Bales}
G.S. Bales and A. Zangwill, Phys. Rev. B {\bf 41}, 5500 (1990).
\bibitem{bray} A.J.\ Bray, Theory of phase-ordering kinetics, Advances
  in Physics, {\bf 43}, 357 (1994). 
\bibitem{harald} H.\ Kallabis and P.\ \v{S}milauer, private
  communications.
\bibitem{gradstein} I.S.\ Gradshteyn and I.M.\ Ryzhik, Tables of
  Integrals, Series and Products, Section {\bf 3.462}, Academic Press
  (1980). 
\bibitem{coniglio} A.\ Coniglio and M.\ Zannetti, Europhys.\ Lett.\
  {\bf 10}, 575 (1989).
\bibitem{pavel}
P. \v{S}milauer and D.D. Vvedensky, 
Phys. Rev. B {\bf 52}, 14263 (1995).
\bibitem{claudio} C.\ Castellano, private communication.

\end{thebibliography}
\end{document}